\documentstyle[12pt]{article}

\catcode`@=11
\def\un#1{\relax\ifmmode\@@underline#1\else
        $\@@underline{\hbox{#1}}$\relax\fi}
\catcode`@=12

\def\a{\alpha}
\def\b{\beta}

\def\d{\delta}

\def\f{\phi}
\def\g{\gamma}
\def\h{\eta}

\def\j{\psi}

\def\l{\lambda}

\def\p{\pi}

\def\L{\Lambda}

\def\ve{\varepsilon}

\def\co{{\cal O}}

\def\bo{{\raise-.5ex\hbox{\large$\Box$}}}               % D'Alembertian
\def\pa{\partial}                                       % curly d
\def\TH{{\raise.2ex\hbox{$\displaystyle \bigodot$}\mskip-4.7mu \llap H \;}}
\def\face{{\raise.2ex\hbox{$\displaystyle \bigodot$}\mskip-2.2mu \llap {$\ddot
        \smile$}}}                                      % happy face
\def\sp#1{{}^{#1}}                              % superscript (unaligned)
\def\Hat#1{\widehat{#1}}                        % big hat
\def\leftrightarrowfill{$\mathsurround=0pt \mathord\leftarrow \mkern-6mu
        \cleaders\hbox{$\mkern-2mu \mathord- \mkern-2mu$}\hfill
        \mkern-6mu \mathord\rightarrow$}
\def\dvec#1{\vbox{\ialign{##\crcr
        \leftrightarrowfill\crcr\noalign{\kern-1pt\nointerlineskip}
        $\hfil\displaystyle{#1}\hfil$\crcr}}}           % <--> accent
\def\frac#1#2{{\textstyle{#1\over\vphantom2\smash{\raise.20ex
        \hbox{$\scriptstyle{#2}$}}}}}                   % fraction
\def\ha{\frac12}                                        % 1/2
\def\sfrac#1#2{{\vphantom1\smash{\lower.5ex\hbox{\small$#1$}}\over
        \vphantom1\smash{\raise.4ex\hbox{\small$#2$}}}} % alternate fraction
\def\bfrac#1#2{{\vphantom1\smash{\lower.5ex\hbox{$#1$}}\over
        \vphantom1\smash{\raise.3ex\hbox{$#2$}}}}       % "
\def\afrac#1#2{{\vphantom1\smash{\lower.5ex\hbox{$#1$}}\over#2}}    % "
\def\[{\lfloor{\hskip 0.35pt}\!\!\!\lceil}
\def\]{\rfloor{\hskip 0.35pt}\!\!\!\rceil}

\def\fracm#1#2{\hbox{\large{${\frac{{#1}}{{#2}}}$}}}
\def\half{{\fracm12}}
\def\ha{\half}

\def\un{\underline}
\def\fracmm#1#2{{{#1}\over{#2}}}

\def\low#1{{\raise -3pt\hbox{${\hskip 0.75pt}\!_{#1}$}}}

\def\Hat#1{\widehat{#1}}

\newskip\humongous \humongous=0pt plus 1000pt minus 1000pt
\def\caja{\mathsurround=0pt}
\def\eqalign#1{\,\vcenter{\openup2\jot \caja
        \ialign{\strut \hfil$\displaystyle{##}$&$
        \displaystyle{{}##}$\hfil\crcr#1\crcr}}\,}
\newif\ifdtup

\def\ref#1{$\sp{#1)}$}

\def\pl#1#2#3{Phys.~Lett.~{\bf {#1}B} (19{#2}) #3}
\def\np#1#2#3{Nucl.~Phys.~{\bf B{#1}} (19{#2}) #3}

\parskip=\medskipamount                 % space between paragraphs (LaTeX)
\thicklines                         % thick straight lines for pictures (LaTeX)

\begin{document}

% =========================== title page ==================================

\thispagestyle{empty}               % no heading or foot on title page (LaTeX)

\def\border{                                            % border
        \setlength{\unitlength}{1mm}
        \newcount\xco
        \newcount\yco
        \xco=-24
        \yco=12
        \begin{picture}(140,0)
        \put(-20,11){\tiny Institut f\"ur Theoretische Physik Universit\"at
Hannover~~ Institut f\"ur Theoretische Physik Universit\"at Hannover~~
Institut f\"ur Theoretische Physik Hannover}
        \put(-20,-241.5){\tiny Institut f\"ur Theoretische Physik Universit\"at
Hannover~~ Institut f\"ur Theoretische Physik Universit\"at Hannover~~
Institut f\"ur Theoretische Physik Hannover}
        \end{picture}
        \par\vskip-8mm}

\def\headpic{                                           % UH heading
        \indent
        \setlength{\unitlength}{.8mm}
        \thinlines
        \par
        \begin{picture}(29,16)
        \put(75,16){\line(1,0){4}}
        \put(80,16){\line(1,0){4}}
        \put(85,16){\line(1,0){4}}
        \put(92,16){\line(1,0){4}}

        \put(85,0){\line(1,0){4}}
        \put(89,8){\line(1,0){3}}
        \put(92,0){\line(1,0){4}}
        \put(85,0){\line(0,1){16}}
        \put(96,0){\line(0,1){16}}
        \put(79,0){\line(0,1){16}}
        \put(80,0){\line(0,1){16}}
        \put(89,0){\line(0,1){16}}
        \put(92,0){\line(0,1){16}}
        \put(79,16){\oval(8,32)[bl]}
        \put(80,16){\oval(8,32)[br]}

        \end{picture}
        \par\vskip-6.5mm
        \thicklines}

\border\headpic {\hbox to\hsize{
\vbox{\noindent ITP--UH--12/94 \\ hep-th/9409017
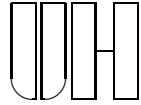
\hfill September 1994}}}

\noindent
\vskip1.3cm
\begin{center}

{\Large\bf        The BRST Charge for the $\hat{D}(2,1;\a)$
                   Non-Linear Quasi-Superconformal Algebra                 }
\footnote{Supported in part by the `Deutsche Forschungsgemeinschaft' and the
NATO grant CRG 930789}\\
\vglue.3in

Sergei V. Ketov \footnote{On leave of absence from:
High Current Electronics Institute of the Russian Academy of Sciences,
Siberian Branch, Akademichesky~4, Tomsk 634055, Russia}

{\it Institut f\"ur Theoretische Physik, Universit\"at Hannover}\\
{\it Appelstra\ss{}e 2, 30167 Hannover, Germany}\\
{\sl ketov@kastor.itp.uni-hannover.de}
\end{center}
\vglue.2in
\begin{center}
{\Large\bf Abstract}
\end{center}
\vglue.2in
\begin{quote}
The quantum BRST charge for the most general, two-dimensional, non-linear,
$N=4$ quasi-superconformal algebra $\hat{D}(1,2;\a)$, whose linearisation
is the so-called `large' $N=4$ superconformal algebra, is constructed. The
$\hat{D}(1,2;\a)$ algebra has
$\Hat{su(2)}_{k^+}\oplus \Hat{su(2)}_{k^-}\oplus\Hat{u(1)}$ Ka\v{c}-Moody
component, and $\a=k^-/k^+$. As a pre-requisite to our construction, we check
the $\hat{D}(1,2;\a)$ Jacobi identities and construct a classical BRST charge.
Then, we analyse the quantum BRST charge nilpotency conditions and find the
only solution, $k^+=k^-=-2$. The $\hat{D}(1,2;1)$ algebra is actually
isomorphic to the $SO(4)$-based Bershadsky-Knizhnik non-linear
quasi-superconformal algebra. We argue about the existence of a new string
theory with (i) the non-linearly realised $N=4$ world-sheet supersymmetry,
(ii) non-unitary matter in a $\hat{D}(1,2;\a)$ representation of $k=-2$ and
$c=-6$, and (iii) negative  `critical dimension'.
\end{quote}

\newpage
\hfuzz=10pt

{\it Introduction}. The critical $N$-extended fermionic string theories
with $N\leq 4$ world-sheet supersymmetries are based on the
{\it two-dimensional} (2d) linear
$N$-extended {\it superconformal algebras} (SCAs) which are gauged \cite{aba}.
The string world-sheet fields usually form a linear $N$-extended superconformal
multiplet coupled to the $N$-extended 2d conformal supergravity fields which
are gauge fields of the $N$-extended SCA. The only known $N=4$ string
theory was constructed by gauging the `small' linear $N=4$ SCA \cite{aba,pn},
and it is of some interest to know
how many different $N=4$ string theories can be constructed at all, despite of
 their apparenly negative `critical dimensions'. The $N=4$ fermionic strings
are relevant in the search for the `universal string theory' \cite{bvu}, and
they are expected to have deep connections with integrable models
\cite{ket,book}.

It has been known for some time that there are {\it two} different
linear $N=4$ SCAs which are (affine verions of) finitely-generated Lie
superalgebras: the so-called `{\it small$\,$}' linear $N=4$ SCA with the
$\Hat{su(2)}$ {\it Ka\v{c}-Moody} (KM) component \cite{aba}, and the so-called
 `{\it large}' linear $N=4$ SCA with the
$\Hat{su(2)}\oplus\Hat{su(2)}\oplus\Hat{U(1)}$ KM component \cite{sch1,stvp}.
Unlike the `small' $N=4$ SCA mentioned above, the `large' $N=4$ SCA has
{\it subcanonical} charges, or `currents' of conformal dimension $1/2$. This
observation already implies that no supergravity or string theory based on the
 `large' $N=4$ SCA exists, because there are no 2d gauge fields which would
correspond to the fermionic charges of dimension $1/2$.~\footnote{In conformal
field theory, `currents' of dimension $1/2$ are just free fermions
\cite{book}.}

However, when a number of world-sheet supersymmetries exceeds two, there are,
in fact, more opportunities to build up new string theories, by using 2d
{\it non-linear} quasi-superconformal algebras which are known to exist for an
arbitrary $N>2$. By an $N$-extended {\it quasi-superconformal algebra} (QSCA)
we mean a graded associative algebra whose contents is restricted to canonical
charges of dimension $2$, $3/2$ and $1$, which (i) contains the Virasoro
subalgebra, and (ii) $N$ real supercurrents of conformal dimension $3/2$,
whose {\it operator product expansion} (OPE) has a stress tensor of dimension
$2$, (iii) satisfies the Jacobi identity, and (iv) has the usual
spin-statistics relation.~\footnote{We exclude from our analysis all kinds of
{\it twisted} (Q)SCAs with unusual relations between spin and statistics (they
are, however, relevant for topological field theory and topological strings
\cite{bv}).} By definition, a QSCA is an `almost' usual SCA, except it may not
be a Lie superalgebra but its OPEs have to be closed on quadratic composites
of the fundamental set of canonical generators. The QSCAs can, therefore, be
considered on equal footing with the $W$ algebras \cite{bs} {\it without},
however, having currents of spin higher than two. Though QSCAs do not belong,
in general, to ordinary (finitely-generated) affine Lie superalgebras, but, so
to say, to infinitely-generated Lie superalgebras, they are still closely
related with finite Lie superalgebras \cite{fl,kac}.

The full classification of QSCAs has been done by Fradkin and Linetsky
\cite{fl}. Their classification is based on the classical results of Ka\v{c}
\cite{kac} about finite simple Lie superalgebras. When $N=4$, the only
different QSCAs are just $su(1,1|2)$ and $\hat{D}(2,1;\a)$. The $su(1,1|2)$
QSCA is, in fact, the `small' $SU(2)$-based linear $N=4$ SCA. The non-linear
$\hat{D}(2,1;\a)$ QSCA was extracted by Goddard and Schwimmer  \cite{gs}
from the `large' linear $N=4$ SCA by factoring out free fermions and boson.
When $\a=1$, i.e. $k^+=k^-\equiv k$, it reduces to the Bershadsky-Knizhnik
$SO(4)$-based quasi-superconformal algebra \cite{kn,be}. The $\hat{D}(2,1;\a)$
QSCA has the non-linear $N=4$ supersymmetry but includes only canonical
charges, which implies the existence of a new $N=4$ conformal supergravity and
a new $N=4$ string theory to be obtained by coupling this supergravity with
 appropriate 2d matter, along the lines of constructing the $W$ gravities and
$W$ strings.
\vglue.2in

{\it The algebra}. Let $J^{a\pm}(z)$ be the internal symmetry currents,
where $a,b,\ldots$ are the adjoint indices of $SU(2)$, and $\pm$ distinguishes
betweeen the two $SU(2)$ factors. We label the four-dimensional fundamental
(vector) representation space of $SO(4)$ by indices $i,j,\ldots\,$. The
self-dual components of the KM currents, $J^{\pm a}(z)$ can be unified into an
antisymmetric tensor $J^{ij}(z)$ in the adjoint of $SO(4)$,
$$J^{ij}(z)=(t^{a-})^{ij}J^{a-}(z) +(t^{a+})^{ij}J^{a+}(z)~,\eqno(1)$$
where the antisymmetric $4\times 4$ matrices $t^{a\pm}$ satisfy the relations
$$\[t^{a\pm},t^{b\pm}\]=-2\ve^{abc}t^{c\pm}~,\quad \[t^{a+},t^{a-}\]=0~,\quad
\{t^{a\pm},t^{b\pm}\}=-2\d^{ab}~.\eqno(2)$$
These matrices can be explicitly represented as
$$(t^{a\pm})^{ij}=\ve^{aij}\pm (\d^i_a\d^j_4-\d^j_a\d^i_4)~,\eqno(3)$$
 and satisfy the identity
$$\sum_a (t^{a\pm})^{ij}(t^{a\pm})^{kl}=\d^{ik}\d^{jl}-\d^{il}\d^{jk}\pm
\ve^{ijkl}~.\eqno(4)$$
The OPEs describing the action of $J^{a\pm}(z)$ read
$$\eqalign{
J^{a\pm}(z)J^{b\pm}(w)~\sim~ & \fracmm{\ve^{abc}J^{c\pm}(w)}{z-w} +
\fracmm{-k^{\pm}\d^{ab}}{2(z-w)^2}~,\cr
J^{a\pm}(z)G^i(w)~\sim~ & \fracmm{{\frac 12}(t^{a\pm})^{ij}G^j(w)}{z-w}~,\cr}
\eqno(5)$$
where {\it two} arbitrary `levels' $k^{\pm}$ for both independent
$\Hat{su(2)}$ KM components have been introduced.

The general ansatz for the OPE of two fermionic supercurrents can be written as
$$
G^i(z)G^j(w)~\sim~  b_1 \fracmm{\d^{ij}}{(z-w)^3} +\fracmm{2T(w)\d^{ij}}{z-w}
 +\ha (b_2 + b_3) \left[\fracmm{J^{ij}(w)}{(z-w)^2} +\fracmm{{\frac 12}\pa
J^{ij}(w)}{z-w}\right]$$
$$ + \ha (b_2- b_3) \ve^{ijkl}\left[\fracmm{J^{kl}(w)}{(z-w)^2} +
\fracmm{{\frac 12}\pa J^{kl}(w)}{z-w}\right]
 +\fracmm{1}{4}b_4 \,\ve^{iklm}\ve^{jkpq}
\fracmm{:J^{lm}J^{pq}:(w)}{z-w}~, \eqno(6a)$$
or, equivalently,
$$\eqalign{
G^i(z)G^j(w)~\sim~ & \fracmm{b_1\d^{ij}}{(z-w)^3} +\fracmm{1}{(z-w)^2}\left[
b_2(t^{a+})^{ij}J^{a+}(w) +b_3(t^{a-})^{ij}J^{a-}(w)\right] \cr
& \fracmm{1}{z-w}\left[ 2T(w)\d^{ij} + \ha\pa\left\{b_2(t^{a+})^{ij}J^{a+}(w)
 +b_3(t^{a-})^{ij}J^{a-}(w)\right\}\right] \cr
& +\fracmm{b_4}{z-w}:\left(t^{a+}J^{a+} -t^{a-}J^{a-}\right)^{(i}{}_k
\left(t^{b+}J^{b+} -t^{b-}J^{b-}\right)^{j)k}:(w)~,\cr}\eqno(6b)$$
where we have used the fact that
$$\ha\ve^{ijkl}\,T^{kl}(z)=\left(t^{a+}\right)^{ij}J^{a+}(z)
-\left(t^{a-}\right)^{ij}J^{a-}(z)~,\eqno(7)$$
as a consequence of eq.~(1).

Demanding associativity of the combinations $TGG$, $JGG$ and $GGG$ determines
the parameters $b_1,\,b_2,\,b_3,\,b_4$, and, hence, all of the QSCA
3- and 4-point `structure constants', {\it viz.}
$$\eqalign{
b_1= \fracmm{4k^+k^-}{k^+ + k^- +2}~,\quad & \quad
b_4= \fracmm{-2}{k^+ + k^- +2}~,\cr
b_2= \fracmm{-4k^-}{k^+ + k^- +2}~,\quad & \quad
b_3= \fracmm{-4k^+}{k^+ + k^- +2}~,\cr}\eqno(8)$$
as well as the central charge,
$$c=\fracmm{6(k^++1)(k^-+1)}{k^++k^-+2}-3~,\eqno(9)$$
in agreement with ref.~\cite{gs}.

We define $\a$-parameter of this $\hat{D}(1,2;\a)$ QSCA as a ratio of its two
KM `levels',
$$\a\equiv  \fracmm{k^-}{k^+}~,\eqno(10)$$
which measures the relative asymmetry between the two $\Hat{su(2)}$ KM
algebras in the whole algebra. When $\a=1$, i.e. $k^-=k^+\equiv k$, the
$\hat{D}(1,2;1)$ QSCA is just the $SO(4)$ Bershadsky-Knihznik QSCA
\cite{kn,be} with the central charge $c=3k$.

In the vector notation, the $\hat{D}(1,2;\a)$ QSCA non-trivial OPEs take
the form
$$
T^{ij}(z)G^k(w)~\sim~ \fracmm{1}{z-w}\left[\d^{ik}G^{j}(w)-\d^{jk}G^{i}(w)
\right]~,\eqno(11a)$$
$$\eqalign{
J^{ij}(z)J^{kl}(w)~\sim~& \fracmm{1}{z-w}\left[\d^{ik}J^{jl}(w)-
\d^{jk}J^{il}(w)+\d^{jl}J^{ik}(w)-\d^{il}J^{jk}(w)\right]\cr
& - \ha(k^+ + k^-)\fracmm{\d^{ik}\d^{jl}-\d^{il}\d^{jk}}{(z-w)^2}
 - \ha(k^+ - k^-)\fracmm{\ve^{ijkl}}{(z-w)^2}~,\cr}\eqno(11b)$$
$$\eqalign{
G^i(z)G^j(w)~\sim~~~ &~\fracmm{4k^+k^-}{(k^+ + k^-+2)}\fracmm{\d^{ij}}{(z-w)^3}
+\fracmm{2T(w)\d^{ij}}{z-w}\cr
&  - \fracmm{k^+ + k^-}{k^+ + k^- +2}
\left[\fracmm{2J^{ij}(w)}{(z-w)^2} +\fracmm{\pa J^{ij}(w)}{z-w}\right]\cr
& + \fracmm{k^+ - k^-}{k^+ + k^- +2}\ve^{ijkl}
\left[\fracmm{2J^{kl}(w)}{(z-w)^2} +
\fracmm{\pa J^{kl}(w)}{z-w}\right] \cr
& - \fracmm{\ve^{iklm}\ve^{jkpq}}{2(k^+ + k^- +2)}
\fracmm{:J^{lm}J^{pq}:(w)}{(z-w)}~.\cr} \eqno(11c)$$

In terms of the Fourier modes of the currents $\co_{\l}$ of dimension
$\l$, defined by $\co_n=\oint (dz/2\p i)\,z^{n+\l-1}\co_{\l}(z)$,  one finds
instead
$$\eqalign{
\[L_m,L_n\]= ~&~ (m-n)L_{m+n}+\fracmm{c}{12}m(m^2-1)\d_{m+n}~,\cr
\[L_m,G^i_r\]= ~&~ ({\frac 12}m-r)G^i_{m+r}~,\qquad
\[L_m,J^{a\pm}_n\]= -nJ^{a\pm}_m~,\cr
\[J^{a\pm}_m,G^i_r\]= ~&~ \ha(t^{a\pm})^{ij}G^j_{m+r}~,\qquad
\[J^{a\pm}_m,J^{b\pm}_n\]=  \ve^{abc}J^{c\pm}_{m+n} -{\frac 12}k^{\pm}m
\d^{ab}\d_{m+n}~,\cr
\{G^i_r,G^j_s\}= ~&~~~ \fracmm{2k^+k^-}{k^++k^-+2}\left(r^2-\ha\right)\d^{ij}
+2\d^{ij}L_{m+n}\cr
& +\fracmm{2}{k^++k^-+2}(s-r)\left[k^-(t^{a+})^{ij}J^{a+}_{r+s}
 + k^+(t^{a-})^{ij}J^{a-}_{r+s}\right]\cr
& -\fracmm{2}{k^++k^-+2}\left(t^{a+}J^{a+} - t^{a-}J^{a-}\right)^{(i}{}_k
\left(t^{b+}J^{b+} -t^{b-}J^{b-}\right)^{j)k}_{r+s}~.\cr}
\eqno(12)$$

Though $\hat{D}(1,2;\a)$ is a non-linear QSCA, it can be turned into a
{\it linear}
SCA by adding some `auxiliary' fields, namely, four free fermions $\j^i(z)$ of
dimension $1/2$, and a free  bosonic current $U(z)$ of
dimension $1$, defining a $\Hat{U(1)}$ KM algebra \cite{gs}. The new fields
have canonical OPEs,
$$\eqalign{
\j^i(z)\j^j(w)~\sim~ & \fracmm{-\d^{ij}}{z-w}~,\cr
U(z)U(w)~\sim~ & \fracmm{-1}{(z-w)^2}~.\cr}\eqno(13)$$
The fermionic fields $\j^i(z)$ transform in a $(2,2)$ representation of
$SU(2)\otimes SU(2)$,
$$J^{a\pm}(z)\j^i(w)~\sim~ \fracmm{{\frac 12}(t^{a\pm})^{ij}\j^j(w)}{z-w}~,
\eqno(14)$$
whereas the singlet $U(1)$-current $U(z)$ can be thought of as derivative of
a free scalar boson, $U(z)=i\pa\f(z)$.

Let us now define the new currents \cite{gs}
$$\eqalign{
T_{\rm tot} = & T -\ha :U^2: - \ha:\pa\j^i\j^i:~,\cr
G^i_{\rm tot} = & G^i - U\j^i
+\fracmm{1}{3\sqrt{2(k^+ +k^-+2)}}\ve^{ijkl}\j^j\j^k\j^l\cr
&  -\sqrt{\fracmm{2}{k^+ +k^-+2}}\,\j^j\left[(t^{a+})^{ji}J^{a+}-
(t^{a-})^{ji}J^{a-}\right]~,\cr
J^{a\pm}_{\rm tot} = & J^{a\pm} +\fracmm{1}{4}(t^{a+})^{ij}\j^i\j^j~,\cr}
\eqno(15)$$
in terms of the initial $\hat{D}(1,2;\a)$ QSCA currents
$T~,\,G^i$ and $J^{a\pm}$. Then the following set of affine generators
$$\left\{ T_{\rm tot}~,\quad  G^i_{\rm tot}~,\quad J^{a\pm}_{\rm tot}~,\quad
\j^i~,\quad U\,\right\}\eqno(16)$$
has closed OPEs among themselves, defining a linear `large' $N=4$ SCA
with the $\Hat{su(2)}\oplus\Hat{su(2)}\oplus\Hat{u(1)}$ KM component!
Explicitly, the non-trivial OPEs of this `large' $N=4$ SCA are given by
({\it cf}~ refs.~\cite{sch1,stvp})
$$\eqalign{
T_{\rm tot}(z)T_{\rm tot}(w)~\sim~ & \fracmm{{\frac 12}(c+3)}{(z-w)^4}
 + \fracmm{2T_{\rm tot}(w)}{(z-w)^2} + \fracmm{\pa T_{\rm tot}(w)}{z-w}~,\cr
T_{\rm tot}(z)\co(w)~\sim~ & \fracmm{h_{\co}\co(w)}{(z-w)^2} +
 \fracmm{\pa \co (w)}{z-w}~,\cr
J_{\rm tot}^{a\pm}(z)J_{\rm tot}^{a\pm}(w)~\sim~ &
\fracmm{\ve^{abc}J_{\rm tot}^{c\pm}(w)}{z-w} - \fracmm{(k^{\pm}+1)\d^{ab}}{2
(z-w)^2}~,\cr
J_{\rm tot}^{a\pm}(z)G^i_{\rm tot}(w)~\sim~ &
\fracmm{{\frac 12}(t^{a\pm})^{ij}G^j_{\rm tot}(w)}{z-w}  \mp
\fracmm{k^{\pm}+1}{\sqrt{2(k^++k^-+2)}}
\fracmm{(t^{a\pm})^{ij}\j^j(w)}{(z-w)^2}~,\cr
G^i_{\rm tot}(z)G^j_{\rm tot}(w)~\sim~ &
\fracmm{{\frac 23}(c+3)\d^{ij}}{(z-w)^3} +\fracmm{2T_{\rm tot}(w)\d^{ij}}{z-w}
 -\fracmm{2}{k^++k^-+2}\left[\fracmm{2}{(z-w)^2}+\fracmm{1}{z-w}\pa_w\right]\cr
& \times \left[(k^-+1)(t^{a+})^{ij}J^{a+}_{\rm tot}(w)+
(k^++1)(t^{a-})^{ij}J^{a-}_{\rm tot}(w)\right]~,\cr
\j^i(z)G^j_{\rm tot}(w)~\sim~ & \fracmm{1}{z-w}\sqrt{
\fracmm{2}{k^++k^-+2}}\left[(t^{a+})^{ij}J^{a+}_{\rm tot}(w)
-(t^{a-})^{ij}J^{a-}_{\rm tot}(w)\right] +\fracmm{U(w)\d^{ij}}{z-w}~,\cr
U(z)G_{\rm tot}^i(w)~\sim~ & \fracmm{\j^i(w)}{(z-w)^2}~,\cr}\eqno(17)$$
where $\co$ stands for the generators $G_{\rm tot},\,J_{\rm tot}$ and $\j$ of
dimension $3/2,\,1$ and $1/2$, respectively, and the $\hat{D}(1,2;\a)$ QSCA
central charge $c$ is given by eq.~(9).~\footnote{Unlike
ref.~\cite{stvp}, we put forward the underlying QSCA structure in our
notation. It is advantageous to express a given algebra in terms of the
smaller number of fundamental charges, whenever it is possible.}

Having restricted ourselves to the (Neveu-Schwarz--type, for definiteness)
Fourier modes
$(L_{\rm tot})_{\pm 1,0}~,~~(G^i_{\rm tot})_{\pm 1/2}$ and
$(J_{\rm tot}^{a\pm})_0\,$, we get a finite-dimensional Lie
superalgebra which is isomorphic to the simple Lie superalgebra $D(1,2;\a)$
from the Ka\v{c} list \cite{kac}. This explains the reason why we use almost
the same (with hat) notation for our affine (infinite-dimensional) QSCA
$\hat{D}(1,2;\a)$ defined by eqs.~(11) or (12). Note that the finite Lie
superalgebra of the `large' $N=4$ SCA in eq.~(17), defining a `linearised'
version of the $\hat{D}(1,2;\a)$ QSCA in eq.~(11), is not simple, but
contains a $U(1)$ piece, in addition to the finite-dimensinal $D(1,2;\a)$
subalgebra. The finite-dimensional simple Lie superalgebras $D(2,1;\a)$ at
various $\a$ values are not, in general, isomorphic to each other (except of
the isomorphism under $\a\to\a^{-1}$, interchanging the two $su(2)$ factors)
\cite{kac}. This is enough to argue about the non-equivalence
(for different $\a$) of the $\hat{D}(1,2;\a)$ QSCAs, which are their affine
generalisations.

It is also worthy to notice that the KM `levels' and the central charge of the
`large' $N=4$ SCA and those of the underlying $\hat{D}(1,2;\a)$ QSCA are
different according to eq.~(16), namely
$$ k^{\pm}_{\rm large} = k^{\pm}+1~,\qquad c_{\rm large} = c+3~,\eqno(18)$$
which is quite obvious because of the new fields introduced. The exceptional
`small' $N=4$ SCA with the $\Hat{su(2)}$ KM component \cite{aba} follows from
eq.~(17) in the limit $\a\to\infty$ or $\a\to 0$, where either $k^-\to\infty$
or $k^+\to\infty$, respectively, and the $\Hat{su(2)}\oplus\Hat{u(1)}$ KM
component decouples from the rest of the algebra. Taking the limit results in
the central charge
$$ c_{\rm small}=6k~,\eqno(19)$$
where $k$ is an arbitrary `level' of the remaining $\Hat{su(2)}$ KM component.
For an arbitrary $\a$, the `large' $N=4$ SCA contains two `small' $N=4$ SCAs
\cite{stvp}.
\vglue.2in

{\it The BRST charge}. Despite of the apparent non-linearity of the
$\hat{D}(1,2;\a)$ QSCA, its quantum BRST charge should be in correspondence
with its classical BRST charge, up to renormalisation. The classical BRST
charge having the vanishing Poisson bracket with itself can, in fact, be
constructed for any algebra of first-class constraints~\cite{ff}. This
provides us with a good {\it ansatz} for the quantum BRST charge we are
looking for. The similar procedure was
 applied to obtain the quantum BRST charge for the non-linear quantum
$W_3$ algebra \cite{tm}, and later generalised to any quadratically non-linear
$W$-type algebra in ref.~\cite{ssvn}. The nilpotency conditions always require
the total (matter + ghosts) central charge to vanish, but also lead to
some more constraints on the QSCA parameters, whose consistency is {\it not}
guaranteed.  This is because the  constraints imposed by the
BRST charge nilpotency condition may be in conflict with the  constraints
dictated by the QSCA Jacobi identities.

The BRST quantisation of the $\hat{D}(1,2;\a)$ QSCA requires the following
ghosts to be introduced:
\begin{itemize}
\item the conformal ghosts ($b,c$), an anticommuting pair of
world-sheet free fermions of conformal dimensions~($2,-1$), respectively;
\item the $N$-extended superconformal ghosts ($\b^i,\g^i$) of conformal
dimensions~($\frac32,-\frac12$), respectively, in the fundamental
representation of $SO(4)\cong SU(2)\otimes SU(2)$;
\item the two pairs of $SU(2)$ ghosts ($\tilde{b}^{a\pm},\tilde{c}^{a\pm}$) of
conformal dimensions~($1,0$), respectively, each one in the adjoint
representation of $SU(2)$.
\end{itemize}

The reparametrisation ghosts
$$b(z)\ =\ \sum_{n\in{\bf Z}} b_n z^{-n-2}~,\qquad
c(z)\ =\ \sum_{n\in{\bf Z}} c_n z^{-n+1}~,\eqno(20)$$
have the following OPE and anticommutation relations:
$$b(z)\ c(w)\ \sim\ \fracmm{1}{z-w}~,
\qquad \{c_m,b_n\}\ =\ \d_{m+n,0}~.\eqno(21)$$

The superconformal ghosts
$$\b^i(z)\ =\ \sum_{r\in{\bf Z}(+1/2)}\b^i_r z^{-r-3/2}~,\qquad
\g^i(z)\ =\ \sum_{r\in{\bf Z}(+1/2)}\g^i_r z^{-r+1/2}~,\eqno(22)$$
satisfy
$$\b^i(z)\ \g^j(w)\ \sim\ \fracmm{-\d^{ij}}{z-w}~, \qquad
 \[\g^i_r,\b^j_s\]\ =\ \d_{r+s,0}~.\eqno(23)$$
An integer or half-integer moding of these generators corresponds to the usual
distinction between the Ramond- and Neveu-Schwarz--type sectors.

Finally, the fermionic $SU(2)\otimes SU(2)$ internal symmetry ghosts
$$\tilde{b}^{a\pm}(z)\ =\ \sum_{n\in{\bf Z}}\tilde{b}^{a\pm}_n z^{-n-1}~,
\qquad
\tilde{c}^{a\pm}(z)\ =\ \sum_{n\in{\bf Z}} \tilde{c}^{a\pm}_n z^{-n}~,
\eqno(24)$$
have
$$\tilde{b}^{a\pm}(z)\ \tilde{c}^{a\pm}(w)\ \sim\ \fracmm{\d^{ab}}{z-w}~,
\qquad
\{\tilde{c}^{a\pm}_m,\tilde{b}^{b\pm}_n\}\ =\ \d^{ab}\d_{m+n,0}~.\eqno(25)$$

In general, given a set of bosonic generators $B_i$ and fermionic generators
$F_{\a}$, satisfying a graded non-linear associative algebra,
$$\eqalign{
\{B_i,B_j\}_{\rm P.B.}=~&~f_{ij}{}^kB_k~,\cr
\{B_i,F_{\a}\}_{\rm P.B.}=~&~f_{i\a}{}^{\b}F_{\b}~,\cr
\{F_{\a},F_{\b}\}_{\rm P.B.}=~&~f_{\a\b}{}^{i}B_{i}+\L_{\a\b}{}^{ij}B_iB_j~,
\cr}\eqno(26)$$
in terms of the graded Poisson (or Dirac) brackets, with some 3-point and
4-point `structure constants', $f_{ij}{}^k$, $f_{i\a}{}^{\b}$, $f_{\a\b}{}^{i}$
 and $\L_{\a\b}{}^{ij}$, respectively, whose values are supposed to be
determined by solving the Jacobi identities, the classial BRST charge $Q$,
satisfying the classical `master equation' $\{Q,Q\}_{\rm P.B.}=0$, is
 given by \cite{ff}
$$\eqalign{
Q = ~&~ c^nB_n + \g^{\a}F_{\a} +\ha f_{ij}{}^kb_kc^jc^i
+ f_{i\a}{}^{\b}\b_{\b}\g^{\a}c^i -\ha f_{\a\b}{}^nb_n\g^{\b}\g^{\a} \cr
&  -\ha \L_{\a\b}{}^{ij}B_ib_j\g^{a}\g^{\b}
-\fracm{1}{24}\L_{\a\b}{}^{ij}\L_{\g\d}{}^{kl}f_{ik}{}^{m}
b_jb_lb_m\g^{\a}\g^{\b}\g^{\g}\g^{\d}~.\cr}\eqno(27)$$
Eq.~(27) may serve as the starting point in a
construction of a quantum BRST charge $Q_{\rm BRST}$ associated with a
quantum non-linear superalgebra. Since we are actually interested in
 quantum QSCAs, we can assume that all operators are
currents, with a holomorphic dependence on $z$ (or, equivalently, with an
additional affine index), and  replace the (graded) Poisson brackets
by (anti)commutators.
In addition, in quantum theory, one must take into account central extensions
and the normal ordering needed for defining products of bosonic generators.
Although no general procedure seems to exist, which would explain how to fully
 `renormalise' the naively quantized (i.e. normally-ordered) charge $Q$ to a
quantum-mechanical operator $Q_{\rm BRST}$, the answer is known for a
particular class of quantum algebras of the $W$-type~\cite{ssvn}. Similarly to
the quantum $W_3$ algebra case considered in ref.~\cite{tm}, the only
non-trivial modification in quantum theory essentially amounts to a {\it
multiplicative} renormalisation of the structure constants $f_{\a\b}{}^i$.
In our case of the $\hat{D}(1,2;\a)$ QSCA, this gives the following
{\it  ansatz}:
$$\eqalign{
Q_{\rm BRST}=~&~ c_{-n}L_n + \g^i_{-r}G^i_r + \tilde{c}^{aA}_{-n}J^{aA}_n
-\ha (m-n)c_{-m}c_{-n}b_{m+n}
+ nc_{-m}\tilde{c}^{aA}_{-n}\tilde{b}^{aA}_{m+n}\cr
& +\left(\fracm{m}{2}-r\right)c_{-m}\b^i_{m+r}\g^i_{-r}
-b_{r+s}\g^i_{-r}\g^i_{-s}
-\ha\tilde{c}^{aA}_{-m}(t^{aA})^{ij}\b^i_{m+r}\g^j_{-r}\cr
& + \h_2 b_2 (r-s) \tilde{b}^{a+}_{r+s}(t^{a+})^{ij}\g^i_{-r}\g^j_{-s}
  + \h_3 b_3 (r-s) \tilde{b}^{a-}_{r+s}(t^{a-})^{ij}\g^i_{-r}\g^j_{-s} \cr
& -\ha \ve^{abc}\tilde{c}^{a+}_{-m}\tilde{c}^{b+}_{-n}\tilde{b}^{c+}_{m+n}
 -\ha \ve^{abc}\tilde{c}^{a-}_{-m}\tilde{c}^{b-}_{-n}\tilde{b}^{c-}_{m+n}
 -\fracmm{1}{2}b_4\L^{ij}_{aAbB}J^{aA}_{r+s+m}\tilde{b}^{bB}_{-m}
\g^i_{-r}\g^j_{-s} \cr
&   -\fracmm{1}{24}b^2_4\L^{ij}_{aAbB}\L^{kl}_{cAdD}
 \ve^{ace}\d_{m+n+p,r+s+t+u}\tilde{b}^{bB}_m\tilde{b}^{dD}_n
(\tilde{b}^{e+}_p+\tilde{b}^{e-}_p)
\g^i_{-r}\g^j_{-s}\g^k_{-t}\g^l_{-u}~,\cr}\eqno(28)$$
where two quantum renormalisation parameters $\h_2$ and $\h_3$ have been
introduced, and $\L^{ij}_{aAbB}$ denote the $\hat{D}(1,2;\a)$ QSCA 4-point
 `structure constants'  $(A=+,-\;$),
$$\L_{aAbB}^{ij}J^{aA}(z)J^{bB}(z)\equiv
 \left(t^{a+}J^{a+}(z)-t^{a-}J^{a-}(z)
\right)^{(i}{}_k \left(t^{b+}J^{b+}(z) -t^{b-}J^{b-}(z)
\right)^{j)k}~.\eqno(29)$$

We find always useful to represent a quantum BRST charge as
$$Q_{\rm BRST}=\oint_0 \fracmm{dz}{2\p i}\,j_{\rm BRST}(z)~,\eqno(30)$$
where the BRST current $j_{\rm BRST}(z)$ is defined {\it modulo} total
derivative.~\footnote{The total derivative can be fixed by requring the
$j_{\rm BRST}(z)$ to transform as a primary field.} In particular, eq.~(28)
can be rewritten as
$$\eqalign{
j_{\rm BRST}(z)=~&~ cT + \g^iG^i + \tilde{c}^{aA}J^{aA} + bc\pa c
- c\tilde{b}^{aA}\pa\tilde{c}^{aA}
-\ha c\g^i\pa\b^i-\fracm{3}{2}c\b^i\pa\g^i  -b\g^i\g^i \cr
&~ -\ha\tilde{c}^{aA}(t^{aA})^{ij}\b^i\g^j
- \left[\h_2 b_2 \tilde{b}^{a+}(t^{a+})^{ij}
  + \h_3 b_3 \tilde{b}^{a-}(t^{a-})^{ij}\right](\g^i\pa\g^j-\g^j\pa\g^i) \cr
& -\ha \ve^{abc}\tilde{c}^{a+}\tilde{c}^{b+}\tilde{b}^{c+}
 -\ha \ve^{abc}\tilde{c}^{a-}\tilde{c}^{b-}\tilde{b}^{c-}
 -\fracmm{1}{2}b_4\L^{ij}_{aAbB}J^{aA}\tilde{b}^{bB}\g^i\g^j \cr
&   -\fracmm{1}{24}b^2_4\L^{ij}_{aAbB}\L^{kl}_{cAdD}
 \ve^{ace}\tilde{b}^{bB}\tilde{b}^{dD}(\tilde{b}^{e+}+\tilde{b}^{e-})
\g^i\g^j\g^k\g^l~.\cr}\eqno(31)$$

The central extensions (anomalies) of the ghost-extended QSCA need not form a
linear supermultiplet, and they actually do not. Therefore, the vanishing of
any anomaly alone does {\it not} automatically mean the vanishing of the
others, unlike in the linear case.

The most tedious part of calculational handwork in computing $Q^2_{\rm BRST}$
can be avoided when using either the Mathematica Package for computing OPEs
\cite{th} or some of the general results in ref.~\cite{ssvn}. In particular,
as was shown in ref.~\cite{ssvn}, quantum renormalisation of the $3$-point
structure constants in the quantum BRST charge should be {\it multiplicative},
whereas the non-linearity $4$-point `structure constants' should {\it not} be
renormalised at all --- the facts already used in the BRST charge ansatz above.
 Most importantly, among the contributions to the $Q_{\rm BRST}^2$, only the
terms {\it quadratic} in the ghosts are relevant. Their vanishing imposes
 the constraints on the central extension coefficients of the QSCA and
simultaneously determines the renormalisation parameter $\h$. The details
can be found in the appendices of ref.~\cite{ssvn}. The same conclusion comes
as a result of straightforward calculation on computer. Therefore, finding out
the nilpotency conditions amounts to calculating only a few terms `by hands',
namely, those which are quadratic in the ghosts. This makes the whole
calculation as simple as that in ordinary string theories based on linear SCAs
 \cite{book}.

The $2$-ghost terms in the $Q_{\rm BRST}^2$ arise from single contractions of
the first three linear (in the ghosts) terms of $Q_{\rm BRST}$ with themselves
and with the next cubic terms of eq.~(28), and from double contractions of
the latter among themselves. They result in the pole contributions to
$j_{\rm BRST}(z)j_{\rm BRST}(w)$, proportional to $(z-w)^{-n}$
with $n=1,2,3,4$. All the residues have to vanish modulo total derivative. We
find
\begin{itemize}
\item from the terms $c(z)c(w)/(z-w)^4\,$:
$$c_{\rm tot}\equiv c+c_{\rm gh}=\left[\fracmm{6(k^++1)(k^-+1)}{k^++k^-+2}-3
\right]+6=0~,\eqno(32a)$$
where the central charge $c$ is given by eq.~(9) and $c_{\rm gh}=+6$;
\item from the terms $\g^i(z)\g^i(w)/(z-w)^3\,$:
$$s_{\rm tot}~\equiv ~b_1 + (b_1)_{\rm gh}~=~ b_1 +\fracmm{3}{2}b_4(k^++k^-)
-6(\h_2b_2+\h_3b_3) +2=0~,\eqno(32b)$$
where the parameters $b_1$, $b_2$ and $b_4$ are given by eq.~(8);
\item  from the terms $\tilde{c}^{a\pm}(z)\tilde{c}^{a\pm}(w)/(z-w)^2\,$:
$$k^{\pm}_{\rm tot}~\equiv ~k^{\pm} + 2 =0~,\eqno(32c)$$
\item from the terms $J^{a\pm}(t^{a\pm})^{ij}\g^i\pa\g^j/(z-w)\,$:
$$-2\h_2b_2-2b_4=-2\h_3b_3-2b_4=0~.\eqno(32d)$$
\end{itemize}

Eq.~(32a) just means the vanishing total central charge, where
the value of $c_{\rm gh}$ is dictated by the standard formula of conformal
field theory \cite{book}
$$\eqalign{
c_{\rm gh}& = 2\sum_{\l} n_{\l}(-1)^{2\l+1}\left(6\l^2-6\l+1\right)\cr
& = 1\times (-26) + 4\times (+11) + 6\times (-2) = +6~,\cr}\eqno(33)$$
 $\l$ is conformal dimension and $n_{\l}$ is a number of the conjugated
ghost pairs: $\l=2,3/2,1$ and $n_{\l}=1,4,6$, respectively. Eq.~(32b) can be
interpreted as the vanishing total supersymmetric anomaly. Since the
supersymmetry is non-linearly realised, this anomaly does not have to vanish
as a consequence of the other equations (32), but, fortunately, it does in our
case. Finally, eqs.~(32c,d) determine $k^{\pm}$ and $\h_{2,3}$.

The only consistent solution to eq.~(32) is
$$k\equiv k^+=k^-=-2~.\eqno(34)$$
This means that the BRST quantisation of the non-linear  $\hat{D}(1,2;\a)$
QSCA can only be consistent if both its $\Hat{su(2)}$ KM components enter
symmetrically, i.e. when this quantum non-linear algebra is actually the
$SO(4)$-based Bershadsky-Knizhnik QSCA of $k=-2$ and $c=3k=-6$. This is to
be compared with the known fact that the quantum BRST charge for the
 `small' $N=4$ SCA, whose all central terms are related and proportional to
central charge, is only nilpotent when $c=-12$.

A connection between the non-linear $SO(4)$-based Bershadsky-Knizhnik QSCA and
the `small' linear $SU(2)$-based SCA exists via the linearisation of the
former into the `large' linear $SU(2)\otimes SU(2)\otimes U(1)$-based SCA and
taking the limit either $k^+\to 0$ or $k^-\to 0$. Since (i) there is no
nilpotent QSCA BRST charge for the case of $k^+\neq k^-$, and (ii) it does not
make sense to gauge and BRST quantise {\it all} the generators of the `large'
$N=4$ linear SCA, there seems to be no direct connection between the
corresponding BRST charges.
\vglue.2in

{\it Conclusion}. In our letter we constructed the quantum BRST charge for
the quantum $\hat{D}(2,1;\a)$ QSCA. It is only nilpotent if $k^+=k^-=-2$, when
the $\hat{D}(2,1;\a)$ QSCA is isomorphic to the $SO(4)$-based
Bershadsky-Knizhnik QSCA.

Gauging the local symmetries of the $SO(4)$-based Bershadsky-Knizhnik QSCA
results in the positive total ghost central charge contribution,
  $c_{\rm gh}=6$. When adding the matter $(\j^i,\f)$ to linearise this algebra,
 one adds $+3$ to the total central charge. In addition, the anomaly-free
solution requires $k=-2<0$. Therefore, there is no way to build an
anomaly-free string theory when using only unitary representations. With a
non-unitary representation of the $SO(4)$-based QSCA of $k=-2$,
one can get the desired anomaly-free matter contribution, $c_{\rm m}=-6$.
Unfortunately, a space-time interpretation and a physical significance of the
construction, if any, then become obscure. Despite of all this, we
 believe that it is worthy to know how many string models, consistent from the
 mathematical point of view, can be constructed. Requiring the existence of
a nilpotent quantum BRST operator, one can construct only two of them having
$N=4$ supersymmetry, either linearly or non-linearly realised.
\vglue.2in

\end{document}